\documentclass[12pt]{article}

\usepackage{amsmath}
\usepackage{amsfonts}
\usepackage{amssymb}
\usepackage{color}

\newcommand{\beq}{\begin{eqnarray}}
\newcommand{\eeq}{\end{eqnarray}}
\newcommand{\R}{\mathbb{R}}

\newcommand{\I}{\mathbb{I}}

\newcommand{\eref}[1]{(\ref{#1})}
\def\Dsl{\,\raise.15ex\hbox{/}\mkern-12.5mu D}
\def\dsl{\,\raise.15ex\hbox{/}\mkern-10.5mu \partial}

\def\ii{\'{\i}}

\begin{document}

\title{\bf  Symplectic gauge fields and dark matter}


\author{
\bf J. Asorey\
\\Department of Physics, University of Illinois at Urbana-Champaign \\ 1110 W. Green St, Urbana, IL 61801, USA\\ \\
\bf M. Asorey\,
\\ Departamento de F\ii sica Te\'orica. Facultad de Ciencias \\ 
Universidad de Zaragoza. 50009 Zaragoza. Spain\\ \\
\bf D. Garc\ii a-\'Alvarez\,\\ Departamento de An\'alisis Econ\'omico \\ Facultad de Econom\ii a y Empresa.\\
 Universidad de Zaragoza. 50005 Zaragoza. Spain}

\maketitle

\begin{abstract}
The dynamics of symplectic gauge fields provides a consistent framework
for fundamental interactions based on
spin three gauge fields.
One remarkable property is that  symplectic gauge fields 
 only  have minimal couplings with gravitational fields and not with any other field of 
 the Standard   Model. 
Interactions  with ordinary matter and radiation can only arise from radiative corrections.  
In spite of the gauge nature of symplectic
fields they acquire a mass by the Coleman-Weinberg mechanism which generates
Higgs-like mass terms where
 the gravitational field  is playing the role of a Higgs field.
Massive symplectic gauge fields
 weakly  interacting with ordinary matter  are  natural candidates for the
dark matter component of the Universe.\end{abstract}

\section{Introduction}

In the Standard Model all fundamental interactions are described by gauge theories.
In the Einstein theory of General Relativity (GR) the gravitational interaction is also formulated in
terms of a gauge field. Although there are significant differences between both
theories, mainly due to the strong connection of GR with the structure
of space-time, the fact that both theories are gauge theories helped to consolidate the gauge 
paradigm where all fundamental interactions are described by gauge fields.

 The search of new physics beyond the Standard Model is supported by
  astrophysical and cosmological evidences  of the existence of a new type of 
invisible matter with unknown interacting properties. The search for new types of interactions
following the gauge principle suggest to explore the possibility of gauge theories with
higher spin \cite{fronsdal, abc}. The pathologies associated to interactions based on 
massless particles with helicities
higher than two  \cite{colemanman}
-\cite{haag} provided an argument to explain why
this kind of  interactions are not observed in Nature. Nevertheless, the challenge is so
interesting that there have been numerous attempts to give a physical meaning to gauge 
theories of higher helicity fields. Free massless fields with arbitrary helicity (or its generalizations)
do exist in any dimension. In fact, Wigner's theory of covariant representations of the Poincar\'e group, 
provides a general  theory of free massless gauge fields  \cite{wignerrep}. Massless fields 
with integer helicity are described by transverse, symmetric  traceless
tensor fields with some equivalence relations which are reminiscent of  gauge transformations 
\cite{fronsdal, abc}. The application of BRST methods to  the consistency analysis 
 of  generalized gauge theories boosted the attempts to extend the analysis of free massless gauge fields 
to interacting theories  from a new viewpoint \cite{bengtssonuno}
-\cite{berends}. The consistency of
the BRST approach requires an infinite tower of higher helicities \cite{bengtssonuno},
\cite{berends}
-\cite{fradkin} in 
close analogy with  string theory. However, even in that case it was impossible to
show the consistency of
the interacting theory \cite{aragone,berendsbis},\cite{bengtssonuno}
-\cite{berends}.

The appearance of  new string  dualities introduced new approaches based on
five-dimensional theories on anti-de Sitter backgrounds \cite{maldacena}
-\cite{Witten}. 
In such a scheme the approach to  higher spin fields acquired a new perspective \cite{vasilievarxiv}
-\cite{vasiliev}. 

In this paper we explore a different  approach to higher spin  gauge fields  based on
a gauge  theory of symplectic fields \cite{invisible}.
In this approach gauge fields are symplectic connections and since their covariant derivatives are  
non-trivial only for fields of spin higher than two,
they are minimally decoupled from the Standard Model physics and only interact with  gravitational fields.
This special characteristic promoted these fields as excellent candidates for the invisible dark matter
component of the Universe.

\section{Symplectic gauge fields}

Let us consider a symplectic form $\omega$ in a four-dimensional space-time\footnote{The theory can be generalized for
arbitrary even dimensional space-times}, i.e. 
 a regular antisymmetric tensor field $\omega_{\mu\nu}=-\omega_{\mu\nu}$ which is closed $d \omega=0$.
The symplectic form $\omega$ can be considered as the antisymmetric component of a generalized space-time metric
in the sense first considered  by R.  Forster (formerly known as R. Bach) and developed by
Schr\"odinger and Einstein in the context of unified field theories. It can also be considered
as a  background electromagnetic field $\omega_{\mu\nu}= \partial_\mu A_\nu-
\partial_\nu A_\mu$ with non-trivial topological density $\epsilon^{\mu_1\mu_2\mu_3\mu_4} \omega_{\mu_1\mu_2}\omega_{\mu_3\mu_4}(x)\neq0$ .

A symplectic gauge field  is by definition a linear  connection 
 which preserves the symplectic form, i.e.  the covariant derivative
 of $\omega$
 \begin{equation}   D_\mu \omega =0\,\end{equation}
 vanishes.
In local coordinates
\begin{equation}
\   \partial_\mu \omega_{\nu \sigma }-\Gamma_{\mu \sigma }^\alpha\, \omega_{\alpha \nu}+\Gamma_{\mu \nu}^\alpha \, \omega_{\alpha
  \sigma}=0,
  \label{simpconn}
  \end{equation}
where $\Gamma_{\mu \sigma }^\alpha$ are the local components of the symplectic 
gauge field. Although, gravitational fields 
are also defined in a similar manner as the linear connections  that preserve the space-time metric symmetric $g$,
the contrast  between both types of fields is very important as we will see below. 

The gauge symmetry is given by space-time transformations which leave the symplectic form invariant ({\it symplectomorphisms}). They are canonical transformations whose  infinitesimal generators   are
given in local coordinates by vector fields of the form 
\begin{equation} \xi_\mu= \partial_\mu \phi,\end{equation}
where $\phi$ is any scalar field. By using canonical transformations it is always possible
to find  local coordinates,   {Darboux\  coordinates}, where $\omega$ becomes a constant form
$$\omega=\begin{pmatrix} 
0 & \I \cr
-\I & 0 
\end{pmatrix}.$$
 In those coordinates, $\partial_\mu \omega=0$  and
\begin{equation}
\Gamma_{\mu \sigma }^\alpha\, \omega_{\alpha \nu}=\Gamma_{\mu \nu}^\alpha \, \omega_{\alpha\sigma}
\label{cuatro}
\end{equation}

If we impose the vanishing of the    torsion  as in the case of a Levi-Civit\` a metric connection, we have  
\begin{equation}
\  \Gamma_{\mu\nu}^\alpha =\Gamma_{\nu\mu}^\alpha.
\label{cinco}
\end{equation}

The  components of a torsionless  symplectic gauge field
in\  Darboux\  coordinates 
\begin{equation}T_{\nu\mu\sigma}=\Gamma_{\mu \nu}^\alpha \, \omega_{\alpha \sigma} \hspace{.7cm}{\  }\end{equation}
define a  3-covariant symmetric tensor
\begin{equation}T_{\nu\mu\sigma}= T_{\mu\nu\sigma}=T_{\nu\sigma\mu}=T_{\mu\sigma\nu}=T_{\sigma \nu\mu}=T_{\sigma\mu\nu }.\end{equation}

Thus, the space of torsionless symplectic gauge fields \cite{Lemlein}
-\cite{gelfand}
can be identified with the space of 3-covariant symmetric tensors. This space of symplectic gauge fields 
is  infinite dimensional  in contrast with the space of Riemannian gauge fields where the Levi-Civit\`a 
connection is unique for any Riemannian metric.

The curvature   tensor $R^\alpha_{\beta \mu\nu}$, 
 \begin{equation}   R^\alpha_{\ \beta \mu\nu}=\partial_\mu \Gamma_{\beta \nu }^\alpha -\partial_\nu \Gamma_{\beta \mu }^\alpha+\Gamma_{ \nu \beta}^\sigma\Gamma_{ \mu \sigma}^\alpha-\Gamma_{ \mu \beta}^\sigma\Gamma_{ \nu \sigma}^\alpha\end{equation} defines by contraction with $\omega$ a (0,4)-tensor
\begin{equation}\   R_{\alpha \beta \mu\nu}= \omega_{\alpha \sigma }R^\sigma_{\ \beta \mu\nu},\end{equation}
with interesting symmetry properties
 \begin{eqnarray*}\  \phantom{\Big(} R_{\alpha \beta \mu\nu}&=& - R_{\alpha \beta \nu \mu}= R_{\beta \alpha\mu\nu},  \\
\  \phantom{\Big(}R_{(\alpha \beta \mu\nu)}&=&R_{\alpha \beta \mu\nu}+R_{\nu \alpha \beta \mu}+R_{\mu\nu \alpha \beta}+R_{\beta \mu\nu\alpha }=0.\end{eqnarray*}
The permutation symmetries of this tensor are characterized by the Young tableau

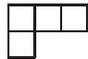
\hspace{5cm}
\begin{picture}(30,20)(0,0)
\put(0,10){\line(1,0){30}}
\put(0,0){\line(1,0){30}}
\put(0,-10){\line(1,0){10}}
\put(0,10){\line(0,-1){20}}
\put(10,10){\line(0,-1){20}}
\put(20,10){\line(0,-1){10}}
\put(30,10){\line(0,-1){10}}
\end{picture}

\bigskip
\noindent
which is in contrast with that of the standard Riemannian  tensor

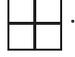
\hspace{5cm}
\begin{picture}(20,20)(0,0)
\put(0,10){\line(1,0){20}}
\put(0,0){\line(1,0){20}}
\put(0,-10){\line(1,0){20}}
\put(0,10){\line(0,-1){20}}
\put(10,10){\line(0,-1){20}}
\put(20,10){\line(0,-1){20}}
\hspace{.8cm}.
\end{picture}
\bigskip

A  symplectic Ricci tensor can also defined by 
\begin{equation}   
\label{ricci}
R_{\beta\nu}=\omega^{\mu\alpha}R_{\alpha \beta \mu\nu},\end{equation}
 and is  symmetric 
\begin{equation}\   R_{\nu \mu}= R_{\mu\nu}, \end{equation}
like the Riemannian Ricci tensor.
However, there is no scalar symplectic curvature because the
contraction of the Ricci tensor with the symplectic form vanishes.

\section{Symplectic Field Theory }

The simplest dynamics for symplectic gauge fields
is defined by the action

\begin{equation}
 \!\!S(\Gamma,\omega)\!=\!\frac1{2{\alpha_0}^2}\!\int\! d^4 x\ R^{\alpha \beta \mu\nu} R_{\alpha \beta \mu\nu} + \frac{{\  \theta}}{32
    \pi^2 }\!\int\! d^4 x \left(R_{\alpha \beta  \mu\nu } R^{ \alpha\beta\mu\nu}\!-\!2 R_{\mu\nu} R^{\mu\nu} \right),
\label{seis}
\end{equation}
which only involves the curvature tensors 
\begin{equation}
{ R}^{\alpha \beta \mu\nu} = \omega^{\alpha\alpha'}\omega^{\beta\beta'}\omega^{\mu\mu'}\omega^{\nu\nu'}R_{\alpha' \beta'
  \mu'\nu'}\qquad { R}^{\mu\nu} = \omega^{\mu\mu'}\omega^{\nu\nu'}R_{ \mu'\nu'}
  \end{equation}
  and the symplectic form $\omega$.
The second term of (\ref{seis}) is proportional to the Pontryagin class of the
 manifold which has a topological meaning and does not contribute to
the classical dynamics.

 The metric independence of (\ref{seis}) implies that the dynamics of the
symplectic fields is completely decoupled from  gravity. 

The action (\ref{seis}) is the most general 
metric independent action of symplectic fields with
quadratic dependence in the curvature tensor \cite{bourgeois}. 
Although  one could  add an extra term
proportional to the square of the Ricci tensor \eref{ricci}, it turns out that such a term is not independent 
of the other two terms of the action \eref{seis}.
Thus, the extra term can be  absorbed by  shifting  the couplings $\alpha_0$ and $\theta$.

The theory is invariant under symplectomorphisms, i.e. canonical transformations.
Symplectic gauge fields, however, transform as
\begin{equation}
T_{\mu\nu\sigma}^\prime=T_{\mu\nu\sigma}+D_\mu D_\nu\partial_\sigma \phi.
\label{siete}
\end{equation}
under symplectomorphisms, where $D_\mu=\partial_\mu+ \Gamma^\sigma_{\mu\nu}$.
The invariance of the action (\ref{seis}) under these transformations implies 
the existence of an infinity of zeromodes.

The field theory governed by (\ref{cinco}) is very interesting from a geometrical  viewpoint  \cite{bourgeois},
but from a quantum field theory perspective it presents many pathologies. The   Cauchy problem is highly  degenerated 
 as it  is pointed out by the existence of
 many   zero modes  in   quadratic terms which are 
not associated to any known gauge symmetry. Apart from the zero modes
associated to the symplectic gauge symmetry (\ref{siete}) there are  nine extra
 zeromodes. The remaining non-null  modes of the quadratic variation of the action on a trivial 
 $T=0$ background are of the form 
$$\textstyle \ \ \ \  \ \ \ \ -\frac1{3}p^2\, , \ \ \ \pm \frac{\sqrt{2}}{3} p^2 \ \  \ \ \ \  \ \ \ \ \ \ \ \    \mathrm {(double\ degenerated)}$$%
$$\textstyle  \frac1{3}  p^2\  \mathrm {}    
, \ \ \  - \frac2{3}  p^2  \mathrm{},     
\ \ \  \pm \frac{{1}}{\sqrt{3}} p^2\  \ \ \ \  \ \ \ \ \mathrm {(non\ degenerated),}    $$
where $p_\mu$ are the momentum of Fourier modes in Darboux coordinates.
Although the eigenvalues of the  quadratic terms of the action are   $SO(4)$ rotation
invariant   the corresponding eigenfunctions $T_{\mu\nu\sigma}$ are not
invariant under  Euclidean or Poincar\'e transformations. This is due to the
background symplectic form  $\omega_{\mu\nu}$ which
introduces a phase space structure in the  space-time which is not compatible
with Euclidean or Poincar\'e symmetries.
Moreover, the quadratic terms  of the action are not 
definite positive   as a consequence of the symplectic structure.
This implies  that the Gaussian projection  defines a  theory with ghost fields 
which is  not   unitary quantum field theory .

  Poincar\'e symmetry can be recovered if we consider a generalization of the action where the symplectic
form $\omega$ becomes a full-fledged dynamical field.  
A  natural choice is to introduce a kinetic term for the symplectic form
$$\frac1{2{\rm e}^2}\int d^4 x\ \omega^{\mu\nu}\omega_{\mu\nu},$$
with $\omega_{\mu\nu}= \partial_\mu A_\nu-
\partial_\nu A_\mu$. But, because of the identity  $ \omega^{\mu\nu}\omega_{\mu\nu}=4$ 
the integrand is  constant  and there is no dynamical content as the trivial motion equations
point out.

The only non-trivial  possibility is to include terms with tensorial contractions which 
involve the space-time metric  (i.e. coupling to gravity).  In this framework it is possible to recover
Poincar\'e invariance  in a Minkowskian metric background.

\section{Interaction with Gravity}

Let us consider  a different  theory of the symplectic gauge fields  interacting with the 
space-time  metric $g$

\begin{equation}S_0(\Gamma,\omega,g)=\frac1{2  
}\int d^4 x\ \sqrt{g} g^{\mu\mu'}g^{\nu\nu'}\omega_{\mu'\nu'}\omega_{\mu\nu}.\end{equation}

Instead of imposing the restriction to the symplectic gauge fields that preserve the symplectic form  $\omega$
\eref{simpconn}, we introduce the constraint in a softer way via a Lagrange multiplier term in the action
\begin{equation}
S_0'(\Gamma,\omega,g)=\frac1{2{\rm \alpha_0}^2}\int d^4 x\ \sqrt{g}\ g^{\gamma\gamma'}g^{\mu\mu'}g^{\nu\nu'}D_{\gamma'}\omega_{\mu'\nu'}D_\gamma\omega_{\mu\nu}.
\label{constr}
\end{equation}
The strong symplectic condition, $D_\gamma\,\omega_{\mu\nu}=0$,
is recovered in the weak coupling limit $\alpha_0\to 0$

The main interaction of symplectic fields with  gravity can be introduced by  contracting 
indices of the curvature tensor with the space-time metric  instead of only using the symplectic form, e.g.
\begin{equation}
  S_1(\Gamma,\omega,g)={{ \alpha}^2}\int d^4x\, \sqrt{\  g}\,
g^{\alpha\alpha'}g^{\beta\beta'}g^{\mu\mu'}g^{\nu\nu'}R_{\alpha' \beta' \mu'\nu'}R_{\alpha \beta \mu\nu}+  \cdots 
\end{equation}
However,  integration over  symplectic forms can generate  new
local terms  in the effective action and the renormalizability condition requires to
consider all possible relevant couplings which do not violate any fundamental gauge symmetry. 
Since the symplectic gauge fields generically do
not preserve the space-time metric
\begin{equation}
D_\sigma\, {g_{\mu\nu}}\neq 0,
\end{equation}
marginally relevant terms of the form  
\begin{equation}  S_1'(\Gamma,\omega,g)=  {{ \alpha_1}^2}\int d^4x\, \sqrt{\  g}\,
|D_\sigma D_\delta \,{ g_{\mu\nu}}|^2+ \cdots\end{equation}
should also be considered because there is no symmetry preventing its 
appearance as radiative corrections.

In summary, one has to include all renormalizable possible independent couplings between gravitational field
and the symplectic gauge field. There are only  six independent types of renormalizable interaction terms
\begin{equation}
  DD{ g}\, DD{ g}, \ 
  D{ g}\, D{g}\, DD{ g}, \ 
  D{ g}\, D{g}\, D{ g}\, D{ g},\
  R\, R, \
  R\, DD{g}, \
  R\, D{g}\, D{ g},
  \label{interaction2}
\end{equation}
because all others can be expressed as linear combinations of these
terms \cite{invisible}.  However, the different contractions of the Lorentz indices
give rise to 78 different  interaction terms involving 78 independent dimensionless 
couplings $\alpha_1,\cdots, \alpha_{78}$: twenty two ($\alpha_{1}\dots \alpha_{22}$) of the type $  DD{ g}\ DD{ g}$, six  ($\alpha_{23}\dots \alpha_{28}$)
 of the type $D{ g}\ D{ g}\ D{ g}\ D{ g}$ and 
 fifty  ($\alpha_{29}\dots \alpha_{78}$) of the type $ D{ g}\ D{ g}\ DD{ g}$. 
 The complete list of these terms is given by equations \eref{22}-\eref{50} in appendix A.

The corresponding theory is renormalizable. In particular,   the effective
action generated by integrating out the symplectic form $\omega$ in the action $S_1$
gets non-trivial contributions to all seventy eight $\alpha$ couplings of symplectic fields with gravity.
In fact, these corrections are logarithmically divergent and the coefficients of 
the corresponding  contributions to the beta functions are displayed in  Table 1 in appendix A.

 We remark that some of the beta functions are positive and some others are negative.
 This means that not all of them will be relevant in the full-fledged quantum theory. However,
 the above calculations have not taken into account the radiative corrections due to
  symplectic gauge field fluctuations. This calculation is beyond the scope of this paper, but it is
  crucial to  elucidate which couplings of the theory are finally relevant. 
 
 The above calculations show that the symplectic field theory is a renormalizable quantum field
 theory, however, the appearance of four order derivative terms in the action introduces some
  ghost components in the symplectic gauge theory. The absence of a larger gauge
 symmetry means that  unitary is not guaranteed.

\section{Symplectic fields and dark matter}
Symplectic gauge fields as  linear connections cannot  interact by minimal couplings with 
scalar fields, because the minimal coupling  in this case reduces to 
$D_\mu\phi=\partial_\mu\phi$. A similar effect
 arises in the interaction with  fermions. The gauge group of symplectic connections is $GL(4, \R)$ and only
 the trivial representation of this group acts on spinors, i.e. there is no analogue of spin connection
 for symplectic gauge fields, then $\Dsl\psi=\dsl\psi$.
 
Thus, the  minimal coupling of  symplectic gauge fields 
to Standard Model particles can only be possible with   gauge particles: the photon or the
intermediate gauge bosons $W^\pm$ and $Z$. However, due to the intrinsic gauge character of these particles
this coupling is not possible. The torsionless character of symplectic  gauge fields is  responsible for the
decoupling also of vector potentials. Indeed,
\begin{equation}F_{\mu\nu}= \partial_\mu A_\nu-\partial_\nu A_\mu + \Gamma_{\mu\nu}^\sigma A_\sigma
- \Gamma_{\nu\mu}^\sigma A_\sigma=\partial_\mu A_\nu-\partial_\nu A_\mu.\end{equation}

Thus,  symplectic gauge fields  cannot  minimally  interact 
with  any  particle of the Standard Model. They can
  only minimally  couple to gravitation, whenever $D_\gamma g_{\mu\nu}\neq 0$. If the corresponding quanta
were massive particles, they will be  natural candidates for
the dark matter component of the Universe and indeed, this is what happens. In the standard $\Lambda$CDM cosmological model  dark matter is usually assumed to be fermionic matter. However, a bosonic component could solve  some dark matter puzzles as we shall discuss below.

However, some non-minimal couplings of symplectic gauge fields with ordinary matter like
$\phantom{,}\phi^\dagger \, \partial_\nu \phi\, D_{\mu}\, g^{\mu\nu}\phantom{,}$, $|\phi|^2\, D_{\mu} D_{\nu}\, g^{\mu\nu}$ or $\bar\psi  \gamma_\nu\, \psi D_\mu g^{\mu\nu}$ 
can arise as radiative corrections. However, the genuine interacting terms of symplectic gauge fields 
with gravitation   \eref{interaction2} are invariant under the signature flip transformation, 
\begin{equation}
 g_{\mu\nu}\to -g_{\mu\nu},
 \end{equation}
 which changes the signature of the metric tensor $g_{\mu\nu}$ from $(1,3)$
 to $(3,1)$. This symmetry acts as a custodial symmetry which prevents the appearance of
 non-minimal coupling between ordinary matter and symplectic gauge fields.
 
Although the coupling of symplectic gauge fields to the symplectic field $\omega$ \eref{constr}  breaks
 the  signature flip symmetry, the effects of such a symmetry breaking only affect   
 the couplings between ordinary matter and  symplectic gauge fields  
 via radiative corrections  at two-loop level.

The breaking of signature flip symmetry also affects the couplings of gravity to symplectic gauge fields
via  one-loop corrections. We have assumed until now
that these couplings are    dimensionless, however, radiative corrections generate  terms of the form
 \begin{equation}  S_1''(\Gamma,\omega,g)=\frac1{2{ \alpha_m}^2}\int d^4x\, \sqrt{\  g}\,
|D_\sigma \,{\  g_{\mu\nu}}|^2+ \cdots.\end{equation}
Since the metric $g$ is not preserved by  symplectic gauge fields nothing prevents
the appearance of these terms with mass square dimension. 
Indeed, such radiative corrections appear in the form 
\begin{eqnarray}\label{efectivamasa}
S_{\text{Higgs}}''&=&\int d^4x\ D_{\gamma_1} g_{\mu_1\nu_1} D_{\gamma_2} g_{\mu_2\nu_2}\ \left(\frac{1}{48}\ 
  g^{\gamma_1\mu_1}g^{\nu_1\mu_2}g^{\gamma_2\nu_2}+\right. \nonumber\\
 &&{}\left.  \frac{5}{16}\ g^{\gamma_1\gamma_2}g^{\mu_1\mu_2}g^{\nu_1\nu_2}-\frac{3}{16}\
  g^{\gamma_1\mu_2}g^{\mu_1\gamma_2}g^{\nu_1\nu_2} \right) I_2,
\end{eqnarray}
of quadratic divergent terms, with
\begin{equation}
I_2= \int \frac{1}{(2\pi)^4}\frac{d^4r}{r^2}.
\label{integralcuaddivergente}
\end{equation}
Thus, such  terms must be included
in the bare action  to ensure  the renormalizability of the theory. Now, in 
a Minkowski background (i.e. $g_{\mu\nu}=\eta_{\mu\nu} $) these terms 
provide a real mass terms for the spin three gauge fields because then
\begin{equation}
S_{\text{Higgs}}''\approx \frac1{2{ \alpha_m}^2}\int d^4x\, \,  \widetilde{T}^{\mu\nu\sigma} T_{\mu\nu\sigma}
\label{mass}
\end{equation}
i.e., although  symplectic gauge fields were in principle related to  massless
particles, they acquire a mass from quantum radiative corrections in
 Minkowski space-time metric backgrounds. The phenomenon is reminiscent of  the Coleman-Weinberg
 mechanism of generation of mass for  conformal scalar electrodynamics. 

The way  symplectic gauge fields $T_{\mu\nu\sigma} $ acquire a mass is also
 reminiscent of the Higgs mechanism with the gravitational field playing the role of the Higgs
field.

 Conversely, the alternative mechanism where symplectic fields condensate into a non-trivial
value and provides a mass terms for the graviton is also possible but not physically 
realistic because a non-trivial expectation value of such a field will break Lorentz 
invariance which is quite unlikely to happen. As a consequence the graviton remains
massless but the symplectic fields become massive.

In a similar manner radiative corrections generate at two-loop level new interacting terms involving
symplectic fields and  Higgs fields of the form
 \begin{equation}
S_{\text{Higgs}}'''(\Gamma,\phi)=\frac1{2{  \alpha_h}^2}\int d^4x\, \sqrt{\  g}\,|\phi|^2
|D_\sigma \,{\  g_{\mu\nu}}|^2 +\cdots,
 \end{equation}
which in  a Minkowski background 
provide  real mass terms for the symplectic gauge fields like in equation \eref{mass}
\begin{equation}
S_{\text{Higgs}}'''\approx \frac{|v|^2}{2{ \alpha_h}^2}\int d^4x\, \,  \widetilde{T}^{\mu\nu\sigma} T_{\mu\nu\sigma},
\label{higgs3}
\end{equation}
where $v=<\!\phi\!> $ is the vacuum expectation value of the Higgs field.
The Higgs contribution to the mass of the symplectic gauge fields \eref{higgs3} is similar 
to the mass terms of the other particles of the Standard Model. The only difference is that  the mass term of
symplectic gauge fields has an extra mass contribution due to   radiative corrections of symplectic fields.

\section{Discussion}
The Standard Model sector of the Universe contains  a large variety of particles. It is then 
envisageable that
 the dark matter sector is also made of more than one type of 
particles. The characteristics of spin 
three massive gauge  particles associated to symplectic gauge fields 
suggest that they are natural candidates as components of dark matter. 
The mass of these gauge particles is only dictated by 
the coupling to gravitation which means that generically it can be 
large enough to provide  a relevant component 
of the cold dark matter. On the other hand, the bosonic character of the new particles
could explain the smooth behavior of the central dark matter density in galaxy halos 
 \cite{{Peebles:1999}}
-\cite{Arbey:2003} and it could give rise to
bosonic condensates which provide interesting scenarios for  dwarf galaxies  \cite{Ji:1994}
-\cite{Arbey:2006}.

Since the only  primary interaction of symplectic gauge fields involves gravitational 
fields  the effect of the new interaction can be mimicked 
by a effective theory of gravitation. The results obtained via integration of  symplectic  gauge fields
yield an effective action which is highly non-local and it  will only become local in the infinite
mass limit of symplectic gauge fields. In that case one  gets  back the standard gravitational action with 
extra $R^2$ terms. However, the physical interpretation of the effective theory is very subtle because
the calculation is highly  dependent on the background space-time metric. 
There are metric backgrounds where the Higgs
mechanism provides a mass to symplectic  gauge fields, and metric backgrounds without such a
mass generating mechanism. In the latter  case the symplectic gauge fields contain massless particles. Thus,
the theory  provides scenarios  which interpolate between hot and cold dark matter scenarios 
depending on the gravitational background. This chameleonic property 
of symplectic gauge fields is very attractive and deserves further exploration.

\bigskip

\section*{Acknowledgments}
We thank  J.M. Mu\~noz-Casta\~neda for discussions. J. A. acknowledges financial support 
from U.S. Department of Energy Grant No. \uppercase{DE-SC}0009932. M.A. has been
partially supported by  Spanish DGIID-DGA Grant No. 2015-E24/2,  Spanish
MINECO Grants No. FPA2012-35453 and No. CPAN-CSD2007-00042 and 
European Cooperation in Science and Technology COST Action MP1405 QSPACE.

\vfill\eject
\appendix
\section{Renormalization} \label{Appendix}

The 78 independent dimensionless couplings of the symplectic gauge fields to gravity 
can be obtained by using Tensorial and FeynCalc packages of Mathematica.

There are three different types of terms: twenty two of the type $DD{ g}\ DD{ g}$,

\noindent\begin{equation}
\begin{array}{lll}
 S'_{22}&=&\displaystyle \int\!\! d^4x  \sqrt{g}\  (D_{\tau_1} D_{\gamma_1}  g_{\mu_1 \nu_1})
\ (D_{\tau_2} D_{\gamma_2}  g_{\mu_2 \nu_2})
\hfill\\
&\Big[&      \displaystyle
 {\alpha}_{1}\ \,  g^{ {\mu_1} {\nu_1}}\  g^{ {\mu_2} {\nu_2}} \ 
 g^{ {\tau_1} {\gamma_1}}\
 g^{ {\tau_2} {\gamma_2}}  +
\displaystyle
 \alpha_{2}\ \,  \ g^{ {\mu_1} {\nu_1}} \ g^{ {\tau_1} {\gamma_1}} \ g^{ {\gamma_2} {\nu_2}}
\ g^{ {\tau_2} {\mu_2}}\\
&+&   \!\!\!  \phantom{[} \displaystyle  \alpha_{3}\ \,  \ g^{ {\mu_1} {\tau_2}} \ g^{ {\mu_2} {\nu_2}} \ g^{ {\nu_1} {\gamma_2}}
\ g^{ {\tau_1} {\gamma_1}}
 +
      \alpha_{4}\ \,  \ g^{ {\mu_1} {\tau_2}} \ g^{ {\nu_1} {\mu_2}} \ g^{ {\tau_1} {\gamma_1}}
\ g^{ {\gamma_2} {\nu_2}}\\
 &+&   \!\!\!    \phantom{[}\displaystyle  \alpha_{5}\ \,  \ g^{ {\mu_1} {\gamma_2}} \ g^{ {\nu_1} {\mu_2}} \ g^{ {\tau_1} {\gamma_1}}
\ g^{ {\tau_2} {\nu_2}}
 +
       \alpha_{6}\ \,  \ g^{ {\mu_1} {\mu_2}} \ g^{ {\nu_1} {\nu_2}} \ g^{ {\tau_1} {\gamma_1}}
\ g^{ {\tau_2} {\gamma_2}}\\
&+&   \!\!\!    \phantom{[}\displaystyle     \alpha_{7}\ \,  \ g^{ {\gamma_1} {\nu_1}} \ g^{ {\tau_1} {\mu_1}} \ g^{ {\gamma_2} {\nu_2}}
\ g^{ {\tau_2} {\mu_2}}
  +    \alpha_{8}\ \,  \ g^{ {\mu_2} {\nu_2}} \ g^{ {\nu_1} {\gamma_2}} \ g^{ {\gamma_1} {\tau_2}}
\ g^{ {\tau_1} {\mu_1}}\\
&+&   \!\!\!   \phantom{[} \displaystyle    \alpha_{9}\ \,  \ g^{ {\nu_1} {\mu_2}} \ g^{ {\gamma_1} {\tau_2}} \ g^{ {\tau_1} {\mu_1}}
\ g^{ {\gamma_2} {\nu_2}}
  +     \alpha_{10}\,   \ g^{ {\nu_1} {\mu_2}} \ g^{ {\gamma_1} {\gamma_2}} \ g^{ {\tau_1} {\mu_1}}
\ g^{ {\tau_2} {\nu_2}}\\
 &+&   \!\!\!    \phantom{[}\displaystyle     \alpha_{11}\,   \ g^{ {\nu_1} {\tau_2}} \ g^{ {\gamma_1} {\mu_2}} \ g^{ {\tau_1} {\mu_1}}
\ g^{ {\gamma_2} {\nu_2}}
  +     \alpha_{12}\,   \ g^{ {\nu_1} {\gamma_2}} \ g^{ {\gamma_1} {\mu_2}} \ g^{ {\tau_1} {\mu_1}}
\ g^{ {\tau_2} {\nu_2}}\\
&+&   \!\!\!   \phantom{[} \displaystyle     \alpha_{13}\,   \ g^{ {\mu_2} {\nu_2}} \ g^{ {\nu_1} {\gamma_2}} \ g^{ {\gamma_1} {\mu_1}}
\ g^{ {\tau_1} {\tau_2}}
  +     \alpha_{14}\,   \ g^{ {\nu_1} {\mu_2}} \ g^{ {\gamma_1} {\mu_1}} \ g^{ {\tau_1} {\tau_2}}
\ g^{ {\gamma_2} {\nu_2}}\\
 &+&   \!\!\!   \phantom{[} \displaystyle     \alpha_{15}\,   \ g^{ {\mu_1} {\nu_1}} \ g^{ {\mu_2} {\nu_2}} \ g^{ {\gamma_1} {\gamma_2}}
\ g^{ {\tau_1} {\tau_2}}
  +    \alpha_{16}\,   \ g^{ {\mu_1} {\mu_2}} \ g^{ {\nu_1} {\nu_2}} \ g^{ {\gamma_1} {\gamma_2}}
\ g^{ {\tau_1} {\tau_2}}\\
  &+&   \!\!\!   \phantom{[} \displaystyle     \alpha_{17}\,   \ g^{ {\mu_1} {\gamma_2}} \ g^{ {\nu_1} {\nu_2}} \ g^{ {\gamma_1} {\mu_2}}
\ g^{ {\tau_1} {\tau_2}}
  +     \alpha_{18}\,   \ g^{ {\mu_1} {\mu_2}} \ g^{ {\nu_1} {\nu_2}} \ g^{ {\gamma_1} {\tau_2}}
\ g^{ {\tau_1} {\gamma_2}}\\
&+&   \!\!\!   \phantom{[} \displaystyle     \alpha_{19}\,   \ g^{ {\mu_1} {\tau_2}} \ g^{ {\nu_1} {\nu_2}} \ g^{ {\gamma_1} {\mu_2}}
\ g^{ {\tau_1} {\gamma_2}}
 +     \alpha_{20}\,   \ g^{ {\nu_1} {\tau_2}} \ g^{ {\gamma_1} {\mu_1}} \ g^{ {\tau_1} {\mu_2}}
\ g^{ {\gamma_2} {\nu_2}}\\
 &+&   \!\!\!   \phantom{[} \displaystyle   \alpha_{21}\,   \ g^{ {\mu_1} {\tau_2}} \ g^{ {\nu_1} {\nu_2}} \ g^{ {\gamma_1} {\gamma_2}}
\ g^{ {\tau_1} {\mu_2}}
  +     {a}_{22}\, \ g^{ {\mu_1} {\tau_2}} g^{ {\nu_1} {\gamma_2}} g^{ {\gamma_1} {\nu_2}}
\ g^{ {\tau_1} {\mu_2}}\Big],
\end{array}
\label{22}
\end{equation}

six  
 of the type $D{ g}\ D{ g}\ D{ g}\ D{ g}$,


\begin{equation}
\begin{array}{lll}
 S'_{6}=\displaystyle \int d^4x &&\!\!\!\!\!\!\!\!\!\!\!\!\!\sqrt{g}\ (D_{\gamma_1}\ g_{\mu_1\nu_1})\ (D_{\gamma_2}\
g_{\mu_2\nu_2})\ (D_{\gamma_3}\ g_{\mu_3\nu_3})\ (D_{\gamma_4}\ g_{\mu_4\nu_4})\\
  &\Big[&\alpha_{23}\        \ g^{\mu_2 {\gamma_3}} \ g^{\nu_1 {\gamma_2}} \ g^{\nu_2\mu_3}
\ g^{\nu_3\mu_4} \ g^{ {\gamma_1}\mu_1} \ g^{ {\gamma_4}\nu_4}\\
&+&
    \alpha_{24}\        \ g^{\mu_2 {\gamma_3}} \ g^{\mu_3 {\gamma_4}} \ g^{\nu_1 {\gamma_2}}
\ g^{\nu_2\mu_4} \ g^{\nu_3\nu_4} \ g^{ {\gamma_1}\mu_1}\\
&+&
    \alpha_{25}\        \ g^{\mu_2\mu_3} \ g^{\nu_1 {\gamma_2}} \ g^{\nu_2\mu_4}
\ g^{ {\gamma_1}\mu_1} \ g^{ {\gamma_3}\nu_3} \ g^{ {\gamma_4}\nu_4}\\
&+&
    \alpha_{26}\        \ g^{\nu_1\mu_2} \ g^{\nu_3\mu_4} \ g^{ {\gamma_1}\mu_1}
\ g^{ {\gamma_2}\nu_2} \ g^{ {\gamma_3}\mu_3} \ g^{ {\gamma_4}\nu_4}\\
&+&
   \alpha_{27}\        \ g^{\nu_1\mu_2} \ g^{\nu_2\mu_3} \ g^{\nu_3\mu_4}
\ g^{ {\gamma_1}\mu_1} \ g^{ {\gamma_2} {\gamma_3}} \ g^{ {\gamma_4}\nu_4}\\
&+&
    \alpha_{28}\        \ g^{\nu_1\mu_2} \ g^{\nu_2\mu_4} \ g^{\nu_3\nu_4}
\ g^{ {\gamma_1}\mu_1} \ g^{ {\gamma_2}\mu_3} \ g^{ {\gamma_3} {\gamma_4}}
\Big],
\end{array}
\label{6}
\end{equation}

 and fifty  of the type $ D{ g}\ D{ g}\ DD{ g}$,
\noindent\begin{equation}
\begin{array}{lllll}
\!\!\! S'_{50}&=&\displaystyle \int d^4x\ \sqrt{g}\  (D_{\gamma_1} g_{\mu_1 \nu_1}) \,(D_{\gamma_2} g_{\mu_2 \nu_2})&\!\!\! \!\!&\!\!\!\!(D_{\tau_3} D_{\gamma_3} g_{\mu_3 \nu_3}) \\
&\Big[&\!\!
      \alpha_{29} \    g^{ {\mu_2} {\tau_3}}  g^{ {\mu_3} {\nu_3}}  g^{ {\nu_1} {\gamma_2}}
g^{ {\nu_2} {\gamma_3}}  g^{ {\gamma_1} {\mu_1}}   
   &\!\!\!\!\!\!\!\!\!\!\!\!\!\!\!    +& \!\!\!\!\!\!\!\!    \alpha_{30}\ \   g^{ {\mu_2} {\tau_3}} g^{ {\nu_1} {\gamma_2}} g^{ {\nu_2} {\mu_3}}
g^{ {\gamma_1} {\mu_1}} g^{ {\gamma_3} {\nu_3}}\\
    & + &  \!\!    \alpha_{31}\ g^{ {\mu_2} {\gamma_3}} g^{ {\nu_1} {\gamma_2}} g^{ {\nu_2} {\mu_3}}
g^{ {\gamma_1} {\mu_1}} g^{ {\tau_3} {\nu_3}}
    &\!\!\!\!\!\!\!\!\!\!\!\!\!\!\!  +& \!\!\!\!\!\!\!\!    \alpha_{32}\ \ g^{ {\mu_2} {\mu_3}} g^{ {\nu_1} {\gamma_2}} g^{ {\nu_2} {\nu_3}}
g^{ {\gamma_1} {\mu_1}} g^{ {\tau_3} {\gamma_3}}
 \\
    & + & \!\!     \alpha_{33}\  g^{ {\mu_3} {\nu_3}} g^{ {\nu_1} {\mu_2}} g^{ {\gamma_1} {\mu_1}}
g^{ {\gamma_2} {\nu_2}} g^{ {\tau_3} {\gamma_3}}
    &\!\!\!\!\!\!\!\!\!\!\!\!\!\!\!  +&  \!\!\!\!\!\!\!\!    \alpha_{34}\ \ g^{ {\nu_1} {\mu_2}} g^{ {\gamma_1} {\mu_1}} g^{ {\gamma_2} {\nu_2}}
g^{ {\gamma_3} {\nu_3}} g^{ {\tau_3} {\mu_3}}\\
    & +& \!\!    \alpha_{35}\ g^{ {\mu_3} {\nu_3}} g^{ {\nu_1} {\mu_2}} g^{ {\nu_2} {\gamma_3}}
g^{ {\gamma_1} {\mu_1}} g^{ {\gamma_2} {\tau_3}}
    & \!\!\!\!\!\!\!\!\!\!\!\!\!\!\! +&\!\!\!\!\!\!\!\!    \alpha_{36}\ \ g^{ {\nu_1} {\mu_2}} g^{ {\nu_2} {\mu_3}} g^{ {\gamma_1} {\mu_1}}
g^{ {\gamma_2} {\tau_3}} g^{ {\gamma_3} {\nu_3}} \\
    & +&\!\!       \alpha_{37}\  g^{ {\nu_1} {\mu_2}} g^{ {\nu_2} {\mu_3}} g^{ {\gamma_1} {\mu_1}}
g^{ {\gamma_2} {\gamma_3}} g^{ {\tau_3} {\nu_3}}
    & \!\!\!\!\!\!\!\!\!\!\!\!\!\!\!+& \!\!\!\!\!\!\!\!    \alpha_{38}\  g^{ {\nu_1} {\mu_2}} g^{ {\nu_2} {\tau_3}} g^{ {\gamma_1} {\mu_1}}
g^{ {\gamma_2} {\mu_3}} g^{ {\gamma_3} {\nu_3}}\\
    & +&\!\!        \alpha_{39}\  g^{ {\nu_1} {\mu_2}} g^{ {\nu_2} {\gamma_3}} g^{ {\gamma_1} {\mu_1}}
g^{ {\gamma_2} {\mu_3}} g^{ {\tau_3} {\nu_3}}
   & \!\!\!\!\!\!\!\!\!\!\!\!\!\!\!  +&\!\!\!\!\!\!\!\!    \alpha_{40}\  g^{ {\nu_1} {\mu_2}} g^{ {\nu_2} {\nu_3}} g^{ {\gamma_1} {\mu_1}}
g^{ {\gamma_2} {\mu_3}} g^{ {\tau_3} {\gamma_3}} \\
    & +& \!\!      \alpha_{41}\  g^{ {\mu_3} {\nu_3}} g^{ {\nu_1} {\tau_3}} g^{ {\nu_2} {\gamma_3}}
g^{ {\gamma_1} {\mu_1}} g^{ {\gamma_2} {\mu_2}}
    &\!\!\!\!\!\!\!\!\!\!\!\!\!\!\!+&\!\!\!\!\!\!\!\!    \alpha_{42}\  g^{ {\nu_1} {\tau_3}} g^{ {\nu_2} {\mu_3}} g^{ {\gamma_1} {\mu_1}}
g^{ {\gamma_2} {\mu_2}} g^{ {\gamma_3} {\nu_3}} \\
    & +& \!\!      \alpha_{43}\  g^{ {\mu_2} {\mu_3}} g^{ {\nu_1} {\tau_3}} g^{ {\nu_2} {\nu_3}}
g^{ {\gamma_1} {\mu_1}} g^{ {\gamma_2} {\gamma_3}}
    &\!\!\!\!\!\!\!\!\!\!\!\!\!\!\! +&\!\!\!\!\!\!\!\!    \alpha_{44}\  g^{ {\mu_2} {\gamma_3}} g^{ {\nu_1} {\tau_3}} g^{ {\nu_2} {\nu_3}}
g^{ {\gamma_1} {\mu_1}} g^{ {\gamma_2} {\mu_3}} \\
    & +&\!\!      \alpha_{45}\  g^{ {\nu_1} {\gamma_3}} g^{ {\nu_2} {\mu_3}} g^{ {\gamma_1} {\mu_1}}
g^{ {\gamma_2} {\mu_2}} g^{ {\tau_3} {\nu_3}}
    &\!\!\!\!\!\!\!\!\!\!\!\!\!\!\!  +& \!\!\!\!\!\!\!\!    \alpha_{46}\  g^{ {\mu_2} {\mu_3}} g^{ {\nu_1} {\gamma_3}} g^{ {\nu_2} {\nu_3}}
g^{ {\gamma_1} {\mu_1}} g^{ {\gamma_2} {\tau_3}}  \\
    & +& \!\!     \alpha_{47}\  g^{ {\mu_2} {\tau_3}} g^{ {\nu_1} {\gamma_3}} g^{ {\nu_2} {\nu_3}}
g^{ {\gamma_1} {\mu_1}} g^{ {\gamma_2} {\mu_3}}
    &\!\!\!\!\!\!\!\!\!\!\!\!\!\!\! +&\!\!\!\!\!\!\!\!    \alpha_{48}\  g^{ {\nu_1} {\mu_3}} g^{ {\nu_2} {\nu_3}} g^{ {\gamma_1} {\mu_1}}
g^{ {\gamma_2} {\mu_2}} g^{ {\tau_3} {\gamma_3}}\\
    & +&\!\!      \alpha_{49}\  g^{ {\mu_2} {\gamma_3}} g^{ {\nu_1} {\mu_3}} g^{ {\nu_2} {\nu_3}}
g^{ {\gamma_1} {\mu_1}} g^{ {\gamma_2} {\tau_3}}
    & \!\!\!\!\!\!\!\!\!\!\!\!\!\!\!+& \!\!\!\!\!\!\!\!    \alpha_{50}\  g^{ {\mu_2} {\tau_3}} g^{ {\nu_1} {\mu_3}} g^{ {\nu_2} {\nu_3}}
g^{ {\gamma_1} {\mu_1}} g^{ {\gamma_2} {\gamma_3}}  \\
    & +&\!\!        \alpha_{51}\  g^{ {\mu_2} {\tau_3}} g^{ {\nu_1} {\mu_3}} g^{ {\nu_2} {\gamma_3}}
g^{ {\gamma_1} {\mu_1}} g^{ {\gamma_2} {\nu_3}}
    &\!\!\!\!\!\!\!\!\!\!\!\!\!\!\!  +&\!\!\!\!\!\!\!\!    \alpha_{52}\  g^{ {\mu_1} {\mu_2}} g^{ {\mu_3} {\nu_3}} g^{ {\nu_1} {\tau_3}}
g^{ {\nu_2} {\gamma_3}} g^{ {\gamma_1} {\gamma_2}} \\
    & +&\!\!       \alpha_{53}\  g^{ {\mu_1} {\mu_2}} g^{ {\nu_1} {\tau_3}} g^{ {\nu_2} {\mu_3}}
g^{ {\gamma_1} {\gamma_2}} g^{ {\gamma_3} {\nu_3}}
    &\!\!\!\!\!\!\!\!\!\!\!\!\!\!\!  +&\!\!\!\!\!\!\!\!    \alpha_{54}\  g^{ {\mu_1} {\mu_2}} g^{ {\nu_1} {\gamma_3}} g^{ {\nu_2} {\mu_3}}
g^{ {\gamma_1} {\gamma_2}} g^{ {\tau_3} {\nu_3}}  \\
& +& \!\!    \alpha_{55}\  g^{ {\mu_1} {\mu_2}} g^{ {\nu_1} {\mu_3}} g^{ {\nu_2} {\nu_3}}
g^{ {\gamma_1} {\gamma_2}} g^{ {\tau_3} {\gamma_3}}
    &\!\!\!\!\!\!\!\!\!\!\!\!\!\!\!  +&\!\!\!\!\!\!\!\!    \alpha_{56} \,\, g^{ {\mu_1} {\tau_3}} g^{ {\mu_2} {\mu_3}} g^{ {\nu_1} {\gamma_3}}
g^{ {\nu_2} {\nu_3}} g^{ {\gamma_1} {\gamma_2}} \\
    & +&\!\!       \alpha_{57}\  g^{ {\mu_1} {\tau_3}} g^{ {\mu_2} {\gamma_3}} g^{ {\nu_1} {\mu_3}}
g^{ {\nu_2} {\nu_3}} g^{ {\gamma_1} {\gamma_2}}
    &\!\!\!\!\!\!\!\!\!\!\!\!\!\!\!  +&\!\!\!\!\!\!\!\!    \alpha_{58}\  g^{ {\mu_1} {\gamma_2}} g^{ {\mu_3} {\nu_3}} g^{ {\nu_1} {\nu_2}}
g^{ {\gamma_1} {\mu_2}} g^{ {\tau_3} {\gamma_3}}\\
    & +&\!\!       \alpha_{59}\  g^{ {\mu_1} {\gamma_2}} g^{ {\nu_1} {\nu_2}} g^{ {\gamma_1} {\mu_2}}
g^{ {\gamma_3} {\nu_3}} g^{ {\tau_3} {\mu_3}}
    & \!\!\!\!\!\!\!\!\!\!\!\!\!\!\! +& \!\!\!\!\!\!\!\!    \alpha_{60}\  g^{ {\mu_1} {\gamma_2}} g^{ {\mu_3} {\nu_3}} g^{ {\nu_1} {\tau_3}}
g^{ {\nu_2} {\gamma_3}} g^{ {\gamma_1} {\mu_2}} \\
    & +& \!\!       \alpha_{61}\  g^{ {\mu_1} {\gamma_2}} g^{ {\nu_1} {\tau_3}} g^{ {\nu_2} {\mu_3}}
g^{ {\gamma_1} {\mu_2}} g^{ {\gamma_3} {\nu_3}}
    & \!\!\!\!\!\!\!\!\!\!\!\!\!\!\!  +&\!\!\!\!\!\!\!\!    \alpha_{62}\  g^{ {\mu_1} {\gamma_2}} g^{ {\nu_1} {\gamma_3}} g^{ {\nu_2} {\mu_3}}
g^{ {\gamma_1} {\mu_2}} g^{ {\tau_3} {\nu_3}}  \\
    & +&\!\!       \alpha_{63}\  g^{ {\mu_1} {\gamma_2}} g^{ {\nu_1} {\mu_3}} g^{ {\nu_2} {\nu_3}}
g^{ {\gamma_1} {\mu_2}} g^{ {\tau_3} {\gamma_3}}
    &\!\!\!\!\!\!\!\!\!\!\!\!\!\!\!  +& \!\!\!\!\!\!\!\!    \alpha_{64}\  g^{ {\mu_1} {\nu_2}} g^{ {\nu_1} {\tau_3}} g^{ {\gamma_1} {\mu_2}}
g^{ {\gamma_2} {\mu_3}} g^{ {\gamma_3} {\nu_3}}\\
    & +&\!\!      \alpha_{65}\  g^{ {\mu_1} {\nu_2}} g^{ {\nu_1} {\gamma_3}} g^{ {\gamma_1} {\mu_2}}
g^{ {\gamma_2} {\mu_3}} g^{ {\tau_3} {\nu_3}}
    &\!\!\!\!\!\!\!\!\!\!\!\!\!\!\!  +& \!\!\!\!\!\!\!\!    \alpha_{66}\  g^{ {\mu_1} {\nu_2}} g^{ {\nu_1} {\mu_3}} g^{ {\gamma_1} {\mu_2}}
g^{ {\gamma_2} {\tau_3}} g^{ {\gamma_3} {\nu_3}}\\
    & +& \!\!    \alpha_{67}\  g^{ {\mu_1} {\nu_2}} g^{ {\nu_1} {\mu_3}} g^{ {\gamma_1} {\mu_2}}
g^{ {\gamma_2} {\gamma_3}} g^{ {\tau_3} {\nu_3}}
    &\!\!\!\!\!\!\!\!\!\!\!\!\!\!\!  +&\!\!\!\!\!\!\!\!    \alpha_{68}\  g^{ {\mu_1} {\nu_2}} g^{ {\nu_1} {\mu_3}} g^{ {\gamma_1} {\mu_2}}
g^{ {\gamma_2} {\nu_3}} g^{ {\tau_3} {\gamma_3}} \\
    & +& \!\!      \alpha_{69}\  g^{ {\mu_1} {\tau_3}} g^{ {\nu_1} {\mu_3}} g^{ {\nu_2} {\nu_3}}
g^{ {\gamma_1} {\mu_2}} g^{ {\gamma_2} {\gamma_3}}
    &\!\!\!\!\!\!\!\!\!\!\!\!\!\!\!  +& \!\!\!\!\!\!\!\!    \alpha_{70}\  g^{ {\mu_1} {\tau_3}} g^{ {\nu_1} {\mu_3}} g^{ {\nu_2} {\gamma_3}}
g^{ {\gamma_1} {\mu_2}} g^{ {\gamma_2} {\nu_3}}\\
    & +& \!\!     \alpha_{71}\  g^{ {\mu_1} {\gamma_3}} g^{ {\nu_1} {\mu_3}} g^{ {\nu_2} {\nu_3}}
g^{ {\gamma_1} {\mu_2}} g^{ {\gamma_2} {\tau_3}}
    &\!\!\!\!\!\!\!\!\!\!\!\!\!\!\!  +&\!\!\!\!\!\!\!\!    \alpha_{72}\  g^{ {\mu_1} {\gamma_3}} g^{ {\nu_1} {\mu_3}} g^{ {\nu_2} {\tau_3}}
g^{ {\gamma_1} {\mu_2}} g^{ {\gamma_2} {\nu_3}}\\
    & +&\!\!     \alpha_{73}\  g^{ {\mu_1} {\mu_3}} g^{ {\nu_1} {\nu_3}} g^{ {\nu_2} {\gamma_3}}
g^{ {\gamma_1} {\mu_2}} g^{ {\gamma_2} {\tau_3}}
    &\!\!\!\!\!\!\!\!\!\!\!\!\!\!\!  +&\!\!\!\!\!\!\!\!    \alpha_{74}\  g^{ {\mu_1} {\tau_3}} g^{ {\mu_2} {\nu_3}} g^{ {\nu_1} {\gamma_3}}
g^{ {\gamma_1} {\nu_2}} g^{ {\gamma_2} {\mu_3}} \\
    & +&\!\!        \alpha_{75}\  g^{ {\mu_1} {\mu_2}} g^{ {\nu_1} {\gamma_3}} g^{ {\nu_2} {\nu_3}}
g^{ {\gamma_1} {\tau_3}} g^{ {\gamma_2} {\mu_3}}
    &\!\!\!\!\!\!\!\!\!\!\!\!\!\!\!  +&\!\!\!\!\!\!\!\!    \alpha_{76}\  g^{ {\mu_1} {\mu_2}} g^{ {\nu_1} {\mu_3}} g^{ {\nu_2} {\nu_3}}
g^{ {\gamma_1} {\tau_3}} g^{ {\gamma_2} {\gamma_3}}\\
    & +&\!\!       \alpha_{77}\  g^{ {\mu_1} {\mu_2}} g^{ {\nu_1} {\mu_3}} g^{ {\nu_2} {\gamma_3}}
g^{ {\gamma_1} {\tau_3}} g^{ {\gamma_2} {\nu_3}}
    &\!\!\!\!\!\!\!\!\!\!\!\!\!\!\!  +&\!\!\!\!\!\!\!\!    \alpha_{78}\  g^{ {\mu_1} {\mu_2}} g^{ {\nu_1} {\tau_3}} g^{ {\nu_2} {\gamma_3}}
g^{ {\gamma_1} {\mu_3}} g^{ {\gamma_2} {\nu_3}}\! 
\Big].
\label{50}
\end{array}
\end{equation}

Integration over the symplectic fields $\omega$ in the action $S_1$
generates logarithmically divergent  contributions to all  
$\alpha$ couplings. The coefficients of these  divergent terms  can be identified with the coefficients of the
beta functions of $\alpha$ couplings  displayed in the  Table 1.

\begin {table}
\begin{center}
\begin{tabular}{l l l l l}
\hline
\hline 
&  & & & \\[-9pt]
  $
  \beta_{{1}} =   -\frac{11}{320} $  &
 $ 
 \beta_{{2}} =\hfill    -\frac{9019}{15360}  $ &
  $
   \beta_{{3}} = \hfill \frac{1103}{6144} $  &
 $ 
 \beta_{{4}} =  \hfill    -\frac{569}{3840} $ &
 $ 
 \beta_{{5}} =    \hfill   -\frac{221}{640} $  
\\[6pt]
\hline  &  & & & \\[-9pt]
    $ 
    \beta_{{6}} =     \hfill   \frac{151}{1536}   $  &
    $ 
    \beta_{{7}} =      \hfill  -\frac{811}{7680}   $  &
   $ 
    \beta_{{8}} =      \hfill -\frac{2173}{7680}   $  &
   $ 
    \beta_{{9}} =    \hfill   \frac{569}{3840}   $  &
    $ 
    \beta_{{10}} =    \hfill   -\frac{481}{3840} $   
\\[6pt]
\hline  &  & &  \\[-9pt]

  $ 
    \beta_{{11}} =    \hfill   -\frac{1}{24}   $  &
    $ 
    \beta_{{12}} =    \hfill   \frac{509}{3840}   $  &
   $ 
    \beta_{{13}} =    \hfill   \frac{1733}{2560}   $  &
   $ 
    \beta_{{14}} =     \hfill  -\frac{1}{32}   $  &
   $ 
    \beta_{{15}} =    \hfill   -\frac{959}{10240}   $  
\\[6pt]
\hline   & & &  \\[-9pt]

   $ 
    \beta_{{16}} =    \hfill   \frac{5}{96}   $  &
   $ 
    \beta_{{17}} =    \hfill   -\frac{5}{96}   $  &
  $ 
    \beta_{{18}} =     \hfill  -\frac{125}{512}   $  &
    $ 
    \beta_{{19}} =    \hfill   \frac{983}{1920}   $  &
    $ 
    \beta_{{20}} =    \hfill   \frac{811}{7680}   $  
\\[6pt]
\hline   & & &  \\[-9pt]
   $ 
    \beta_{{21}} =   \hfill    \frac{161}{3840}   $  &
   $ 
    \beta_{{22}} =    \hfill   -\frac{143}{1280}   $  &
     $ 
    \beta_{{23}} =     \hfill \frac{353}{1024}   $  &
     $ 
    \beta_{{24}} =     \hfill -\frac{77}{7680}   $  &
    $ 
    \beta_{{25}} =     \hfill -\frac{6691}{30720}   $  
\\[6pt]    \hline   & & & & \\[-9pt]

     $ 
    \beta_{{26}} =     \hfill -\frac{1}{80}   $  &
   $ 
    \beta_{{27}} =     \hfill -\frac{601}{1280}   $  &
     $ 
    \beta_{{28}} =     \hfill \frac{1981}{1920}   $ &
   
    $ 
    \beta_{{29}} =     \hfill  \frac{3299}{10240}   $  &
     $ 
    \beta_{{30}} =    \hfill   \frac{121}{480} $ 
       \\[6pt]
\hline &  & &  \\[-9pt]
  $ 
    \beta_{{31}} =     \hfill  \frac{8977}{15360}   $  &
     $ 
    \beta_{{32}} =     \hfill  \frac{151}{1280}   $  &
     $ 
    \beta_{{33}} =     \hfill  \frac{2283}{5120}   $  &
     $ 
    \beta_{{34}} =     \hfill  \frac{1083}{2560}   $  &
     $ 
    \beta_{{35}} =    \hfill   \frac{293}{3840}   $  
 
\\[6pt]    \hline   & & &  \\[-9pt]

     $ 
    \beta_{{36}} =     \hfill  -\frac{13}{960}   $  &
     $ 
    \beta_{{37}} =     \hfill  -\frac{1447}{3840}   $  &
     $ 
    \beta_{{38}} =    \hfill   \frac{4909}{15360} $   &
    $ 
    \beta_{{39}} =     \hfill -\frac{15437}{30720}   $  &
     $ 
    \beta_{{40}} =     \hfill -\frac{169}{960}   $  
\\[6pt]    \hline   & & &  \\[-9pt]
     $ 
    \beta_{{41}} =    \hfill   -\frac{1807}{15360}   $  &
    $ 
    \beta_{{42}} =     \hfill  -\frac{95}{256}   $  &
    $ 
    \beta_{{43}} =     \hfill \frac{187}{7680}   $  &
     $ 
    \beta_{{44}} =     \hfill \frac{121}{160}   $ &
    $ 
    \beta_{{45}} =     \hfill -\frac{8459}{7680}   $  
\\[6pt]   
 \hline   & & & &  \\[-9pt]
     $ 
    \beta_{{46}} =     \hfill -\frac{101}{384}   $  &
        $ 
    \beta_{{47}} =     \hfill -\frac{89}{96}   $  &
     $ 
    \beta_{{48}} =     \hfill -\frac{769}{7680}   $  &
     $ 
    \beta_{{49}} =     \hfill \frac{167}{15360}   $  &
    $ 
    \beta_{{50}} =     \hfill \frac{8647}{30720}   $  

%
\\[6pt]    \hline   & & & & \\[-9pt]
    $ 
    \beta_{{51}} =     \hfill \frac{349}{3840}   $  &
     $ 
    \beta_{{52}} =     \hfill \frac{2449}{3840}   $  &
     $ 
    \beta_{{53}} =     \hfill \frac{3323}{15360}   $  &
     $ 
    \beta_{{54}} =     \hfill -\frac{6377}{15360}   $  &
    $ 
    \beta_{{55}} =     \hfill \frac{271}{320}   $  
\\[6pt]    \hline   & & & & \\[-9pt]
     $ 
    \beta_{{56}} =     \hfill \frac{1921}{3072}   $ & 
     $ 
    \beta_{{57}} =     \hfill -\frac{3407}{15360}   $  &
    $ 
    \beta_{{58}} =     \hfill -\frac{457}{768}   $ &
    $ 
    \beta_{{59}} =     \hfill \frac{629}{15360}   $  &
     $ 
    \beta_{{60}} =     \hfill -\frac{57}{512}   $  

\\[6pt]    \hline   & & & & \\[-9pt]
    $ 
    \beta_{{61}} =     \hfill \frac{61}{640}   $  &
     $ 
    \beta_{{62}} =     \hfill \frac{1453}{7680}   $  &
    $ 
    \beta_{{63}} =     \hfill -\frac{695}{3072}   $  &
     $ 
    \beta_{{64}} =     \hfill -\frac{55}{64}   $ &
     $ 
    \beta_{{65}} =     \hfill -\frac{33}{2560}   $  

\\[6pt]    \hline   & & & & \\[-9pt]
     $ 
    \beta_{{66}} =     \hfill -\frac{513}{5120}   $  &
    $ 
    \beta_{{67}} =     \hfill \frac{73}{960}   $  &
     $ 
    \beta_{{68}} =     \hfill -\frac{351}{2560}   $  &
     $ 
    \beta_{{69}} =     \hfill \frac{1}{2}   $  &
    $ 
    \beta_{{70}} =     \hfill \frac{203}{960}   $  
\\[6pt]    \hline   & & & & \\[-9pt]
    $ 
    \beta_{{71}} =     \hfill \frac{7253}{15360}   $ &
     $ 
    \beta_{{72}} =     \hfill \frac{1753}{15360}   $  &
     $ 
    \beta_{{73}} =     \hfill -\frac{15}{64}   $  &
    $ 
    \beta_{{74}} =     \hfill -\frac{6607}{15360}   $  &
     $ 
    \beta_{{75}} =     \hfill -\frac{1417}{1920}   $  
\\[6pt]    \hline   & & & & \\[-9pt]
    $ 
    \beta_{{76}} =     \hfill \frac{2603}{7680}   $  &
    $ 
    \beta_{{77}} =     \hfill -\frac{329}{640}   $  &
    $ 
    \beta_{{78}} =     \hfill \frac{1309}{2560}   $ &
\\[6pt]    \hline \hline
\end{tabular}
\end{center}
\caption {Beta function coefficients of gravitational $\alpha$ couplings  of symplectic gauge fields}
\end{table}

\vskip 3cm 

\bigskip
The fact that  no new couplings are generated by one loop diagrams
points out the renormalizable character of the theory.
The couplings whose  beta function coefficients are listed in Table 1  
are the only dimensionless renormalized couplings of the theory.

\bibliographystyle{aipproc}   

\bibliography{sample}

\end{document}